\documentclass[10pt,a4paper]{article}
\usepackage{graphicx}
\usepackage{natbib}

\def\ee{\end{equation}}
\def\be{\begin{equation}}
\newcommand{\beqn}{\begin{eqnarray}}
\newcommand{\eeqn}{\end{eqnarray}}

\begin{document}

\author{S. Bonazzola$^1$, L. Villain$^{1,2}$ and M. Bejger$^{1,3}$\\
\footnotesize $^1$LUTh, CNRS, Observatoire de Paris, F-92195 Meudon Cedex, 
France\\ \footnotesize
$^2$DFA, Universitat d'Alacant, 
Ap. Correus 99, 03080 Alacant, Spain\\
\footnotesize $^3$N. Copernicus Astronomical Center, PAN, Bartycka 18, 
PL-00-716 Warszawa, Poland}
\title{MHD of rotating compact stars with spectral methods: description of the algorithm and tests}
\date{}
\maketitle
\abstract{A flexible spectral code for the study of general relativistic
  magnetohydrodynamics is presented. Aiming at investigating the physics of
  slowly rotating magnetized compact stars, this new code makes use of various
  physically motivated approximations. Among them, the relativistic anelastic
  approximation is a key ingredient of the current version of the code. In
  this article, we mainly outline the method, putting emphasis on algorithmic
  techniques that enable to benefit as much as possible of the non-dissipative
  character of spectral methods, showing also a potential astrophysical
  application and providing a few illustrative tests.}

\section{Introduction and motivation}
\label{intro}

During the last decades, Astrophysics has seen fast progress in its
observational techniques, and a very wide flow of data is now coming
from several orbital and ground-based detectors. This huge amount of
information is a challenge not only to the data analysis community,
but also to theorists, as accurate numerical models become more and
more required to explain various astrophysical observations and to put
to the test more qualitative theoretical predictions. Among the
processes that are the most interesting and complicated from both
physical and numerical points of view, those related with strong
gravitational and/or magnetic fields play crucial roles inside or in
the vicinity of compact objects. Hence, they are involved in the
description of numerous phenomena, such as are massive stars core
collapses, emergence of jets from active galaxies, emission of
gamma-ray bursts (GRB) or of gravitational waves (GW), merger of
binary neutron stars (NS) or just evolution of isolated NS. As a
consequence, various codes for relativistic MagnetoHydroDynamics (MHD)
have been developed [e.g. \cite{koi99}, \cite{dvh03}, \cite{gam03},
\cite{kom04}, \cite{shi05}, \cite{du05}, \cite{ant06}, \cite{miz06}],
with different scientific purposes, and therefore distinct choices in
the physical approximations, in the writing of the equations and in
the numerical implementation.\\

In the following, we present a new relativistic MHD code, mainly describing
the methods and some basic tests. This code is tridimensional and based on
spectral methods, quite similar to the relativistic hydrodynamical code
introduced in \cite{VB2002} (hereafter referred to as Paper I) and
\cite{vbh05} (Paper II). The new code was designed to be flexible and to
fulfill several demands, the first of them being to solve MHD equations in the
fixed curved space-time associated with the steady-state configuration of a
slowly rotating NS ({\it i.e.} it currently works in the Cowling
approximation). Those equations are solved in the whole interior of the star
(including the origin), using {\it spherical vector components} and {\it
  spherical coordinates}, which is almost necessary to easily implement
various boundary conditions (BC), and hence to be able to deal with a wide
spectrum of physical situations. Another requirement for exploring numerous
physical phenomena is that of the code being robust and stable in the sense of
allowing to follow the evolutionary tracks during many rotational periods. In
such a way, it can also be used to study the stability of equilibrium
configurations of rotating magnetized NSs. As we shall illustrate, this is
made possible by the non-dissipative character of spectral methods and by some
cautions taken in the algorithm, such as
having the numerical magnetic field that remains divergence-free.\\

The article is divided in several sections: in Sect.~\ref{eq}, we
summarized the typical orders of magnitude and timescales involved in
the physical situations that we shall study in the future, using the
resulting approximations to write the relativistic equations of motion
in a Newtonian-like way. Then, the numerical method is described in
Sect.~\ref{method}. We discuss some possible applications in 
Sect.~\ref{non_dissipative_mhd}. Finally, Sect.~\ref{tests} contains tests
performed, whereas final conclusions are gathered in
Sect.~\ref{conclusions}.

\section{Timescales and equations of motion}
\label{eq}

We shall briefly remind orders of magnitude and timescales for the
type of problems to be addressed with the code: We are interested
in the MHD of old but not too cold (not superfluid) compact stars,
which have radii of some $10^6~{\rm cm}$, mean baryonic densities
$n\,\sim\,10^{14}~{\rm g/cm^3}$, and magnetic fields of $10^{12}$ to
$10^{16}$ gauss. We consider only compact stars whose density does not
vanish at the surface (the so-called bare strange stars) or which have a
rigid crust. In the latter case the domain of integration is limited
to the liquid part of the star, completely liquid stars (just born
very hot NSs) being much more difficult to treat since BC of wind
solutions have to be imposed\footnote{This kind of problem can be
treated with two different numerical techniques: A finite difference
method [like the one developed by the Valencia group, see for instance
\cite{ant06}] during the in-falling and cooling phase, then the
present algorithm once the crust is formed. On NSs winds, see also
\cite{buc06}.}.\\

Most of known isolated rotating NSs do not have rotational frequency larger
than 300 Hz ($\sim\,\Omega =$ 1880 rad/s), which gives a period
$\tau_{rot}\,=\,3.3\,10^{-3}$ s. Retaining this typical value implies that to
keep only terms linear in $\Omega$ is a good approximation, be it for the
relativistic Euler equation or the background metric. Using also the conformal
flatness approximation, the latter is written as in Papers I and II in a
$(3+1)$-like isotropic way (in $c=1$ units) \be \label{met} ds^2\,=\,-\,N^2
dt^2\,-\,2\,r^2\,\sin(\theta)^2\,a^2\,N_3\,d\phi\,dt\,+a^2\,dl^2\,, \ee where
$dl^2$ is the Euclidean infinitesimal 3-length, while $r$ and $\theta$ are
spherical coordinates, with the lapse $N$, the conformal
factor $a$, and the shift $N_3$ that only depend on $r$.\\

Another important velocity for magnetized NSs is the Alfv{\'e}n velocity
defined as $c_a\,\sim\,B/\sqrt{4\pi n}\,\sim\,10^4\,B_{12}$ cm/s [with
$B_{12}\,=\,B/(10^{12} \textrm{gauss})$], leading to the Alfv{\'e}n crossing
time $\tau_{Al} = 10^2 \, B^{-1}_{12}$ s (and to $\nu_{Al}\,=\,10^{-2}\,
B_{12}$ Hz). This time and the period are much larger than the acoustic crossing
time $\tau_{ac}$, whose value comes from the sound velocity that, in compact
stars, is approximately equal to one third of the light velocity ($c_s \simeq
0.3\,$c $\to\,\tau_{ac} \sim 3.3\,10^{-5}$ s, with corresponding frequency:
$\nu_{ac} \sim\, 3.\,10^5$ Hz). As $\tau_{ac}$ is much shorter than the other
characteristic times, we shall filter acoustic waves through the so-called
relativistic anelastic approximation introduced in Paper I: \be\label{eq:anel}
\frac{1}{\sqrt{e}}\,\partial_i(\sqrt{e}\,{\cal P}^i)\,=\,0\,, \ee with $e$ the
determinant of the Euclidean 3-metric and ${\cal
  P}^i\,\equiv\,n\,N\,a^3\,U^i$, where $U^i$ is the spatial part of the fluid
4-velocity ($i\,\in\,[1,3]$). Notice that contrary to what was done in Papers
I and II, this relation is not necessarily linearized here, for we shall
also use the divergence-free character of the $P^i$ 3-vector as a key feature
for the algorithmic (see Sect. \ref{method}). Moreover, we shall already
mention that the momentum density ${\cal P}$ (which is the $1$-form associated
to $\vec{{\cal P}}$) is the main variable retained in the hydrodynamical part
of the new code, while in previous works (Papers I and II), this variable was
the velocity $\vec{V}$. The reason of the previous choice was that the
vanishing matter density at the surface of the star (at least in Paper I),
would not have made possible in general to recover the velocity from the
momentum density\footnote{The division of $\vec{\cal{P}}$ by the matter
  density $n$ is numerically possible only for stiff equation of state
  ($\Gamma \ge
  2)$.}.\\

As the star can be described as composed of
a single neutral fluid (see Paper II), we follow \cite{w72} and use both the form of the metric (\ref{met}) and the
anelastic approximation (\ref{eq:anel}), to write Euler equation in the
Newtonian-like way
\beqn \label{MHDP} \partial_t {\cal
P}_i\,+\,\frac{n\,N}{f}\,\overrightarrow{\nabla}_j \left(\frac{f}{n}
U^j {\cal
P}_i\right)\,+\,\frac{a\,n\,N^2}{f}\,\overrightarrow{\nabla}_i
P\,=\nonumber \\\,\frac{{\cal P}_\mu\,{\cal P}_\nu}{2\,a^2\,N\,{\cal
P}^0}\,\overrightarrow{\nabla}_i
g_{\mu\nu}\,+\,\frac{a^3\,N^2\,n}{f}\,{\cal F}_{i\mu}\,{\cal J}^\mu\,,
\eeqn where $f\equiv\,\rho\,+\,P$, whereas ${\cal F}_{i\mu}\,{\cal
J}^\mu$ is the Lorentz force.\\

Mainly due to the frame-dragging effect, to curvature of space-time and
to the possible non-ideality of the plasma, the Lorentz force contains numerous
terms that we shall not all make explicit here. We shall just write it
in such a way to make more visible the equivalent of the Newtonian
force, which will be used in Sect. \ref{tests} to proceed to basic tests of
the code \be {\cal F}_{i\mu}\,{\cal
J}^\mu\,\sim\,\frac{Z}{4\,\pi}\,\left(\vec{\nabla}\wedge\,\vec{B}\right)\wedge\,\vec{B}\,+\,RT\,,
\ee where $Z$ is a metric factor, $\vec{B}$ a magnetic field-like
vector and $RT$ encompasses all other Relativistic Terms (related with
frame-dragging, non-ideality of the conductivity, {\it
etc.}). $\vec{B}$ is defined as \be
B^i\,=\,\frac{1}{2}\,\eta^{ijk}\,F_{jk}\,, \ee with $\eta^{ijk}$ the
tridimensional Euclidean Levi-Civita volume form and $F_{jk}$ the
Maxwell electromagnetic tensor. Its evolution is ruled by the so-called
induction equation, which comes from relativistic Ohm's law\footnote{See for instance \cite{lich67}.} applied to a neutral plasma
\be
\kappa\,J_\mu\,=\,F_{\mu\nu}U^\nu\,,
\ee
where $\kappa$ is the conductivity\footnote{Notice that the decay time of magnetic field in NS is,
  as the viscous time, much larger than the dynamical time under consideration
  here. See for instance \cite{gr92}.}, and from the homogeneous Maxwell equations
\be
\partial_\alpha\,F_{\beta\gamma}\,+\,\partial_\beta\,F_{\gamma\alpha}\,+\,\partial_\gamma\,F_{\alpha\beta}=0.
\ee

The relativistic induction equation for $\vec{B}$ is written in a Newtonian-like way as
\be
\label{Cns} \partial_t \vec{B} + \vec{\nabla} \wedge ( \vec{B} \wedge
\vec{V})\,+\,\vec{\nabla} \wedge (\kappa \vec{{\cal J}})\,=\,0\,,\ee
where $V^i=U^i/U^0$ and ${\cal J}^i=N\,J^i$. Notice that due to the
frame-dragging effect, $J^i$ is not directly related to the curl of
$\vec{B}$ by the inhomogeneous Maxwell equation, but we shall not
enter more into the detail here, and we just mention that while
testing the code, we used the Newtonian expression of Faraday's
equation in the MHD approximation
\be
\frac{1}{4\,\pi}\,\vec{J}\,=\,\vec{\nabla}\,\wedge\,\vec{B}.
\ee

The main reason behind that approximation is that it does not change
the algorithmic but allows to verify the conservation of energy (see
Sect.\ref{tests}) and consequently the whole mechanics of the code and
its non-dissipative character. Furthermore, we would like to insist on
the fact that what we present here is the simplest version of the
code, containing the minimal number of physical assumptions. Thanks to
the required flexibility, more physical situations can be easily
considered (for example strong stratification, see further), by
following the same strategy as going from Paper I to Paper
II. However, notice that when treating non-linear problems, we will
need some kind of artificial viscosity in order to smooth the
solutions. Yet, the main difference between spectral and finite
difference scheme codes is that with spectral codes, artificial
viscosity and resistivity are put as differential operators, implying
that we perfectly know their behaviour, while they also help to impose
boundary conditions. Since for linearized\footnote{Notice that what we
call \emph{the linear case} is a linearization around the rigid
rotation solution, implying equations of motion very similar to the
ones described in Paper I and II.} problems, we do not need, in
principle, dissipative terms, specific problems can arise due to the
fact that instabilities have much more possibilities to
develop. Consequently, the linearized non-dissipative version of the
code is still in progress, even if some tests are already provided
here (see Sect. \ref{non_dissipative_mhd}), showing that the global strategy works well.\\

\section{The method}
\label{method}

\subsection{Implementation of the anelastic approximation}

As in Papers I and II, the solving of the equations (linear or not)
relies on two main ingredients:
\begin{itemize}
\item the Helmholtz theorem, that is to say the decomposition of vectors into divergence-free components and potential parts,
\item the use of so-called angular potentials.
\end{itemize}

However, working with as variable the momentum makes the
implementation of the anelastic approximation quite different, and we
proceed by considering the equation of motion (Eq.~\ref{MHDP}) written
in the following way: \be\label{MHDP2} \partial_t {\cal
P}_i+\partial_i \tilde{P} =S_i\,,\ee where $\tilde{P}$ is a kind of
geometrically modified pressure ($\tilde{P}\equiv\,a\,N^2\,\Pi$ with
$d \Pi\equiv n dP/f$), while the term $S_i$
represents all other terms appearing in Eq.~(\ref{MHDP}) (non-linear included) and which are
considered as a known source. Applied to this known source term $S_i$,
Helmholtz theorem gives a divergence-free component $\tilde{S}_i$ and
a potential part $\partial_i \Psi$:

 \be\label{dec} S_i=\tilde{S}_i + \partial_i \Psi. \ee
 
 Note that the above decomposition is not unique, since one can add an
 arbitrary harmonic function $\Psi_A$ to the potential $\Psi$ (or equivalently
 subtract its gradient to $\tilde{S}_i$) without altering the divergence of
 $S_i$. As we shall see in the following, this degree of freedom is used to
 impose an arbitrary BC for $\vec{\cal P}$ (for example ${\cal P}_r=0$). Due to the anelastic
 approximation and the exact compensation that is required between
 $\partial_i \tilde{P}$ and $\partial_i \Psi$, the equation to be solved reads
\be\label{MHDa} 
\partial_t {\cal P}_{i}=\tilde{S}_{i}.
\ee

If the non-linear equations are considered, a viscosity $\zeta$
(supposed to be constant) is added, and we have
\be\label{MHDav}
\partial_t {\cal P}_i + \zeta \triangle {\cal P}_i=\tilde{S}_i 
\ee

Once $\tilde{P}$ is obtained ($\tilde{P}=\Psi$), the new source term for the next time
step can be calculated (in the barotropic case) by using the equation
of state and the frozen metric. But before entering more into the detail of solving such equations
(the induction equation being technically identical), we shall now briefly recap on angular potentials and their use in spherical geometry.

\subsection{Vector angular potentials} 

As mentioned in Sect.~\ref{intro}, we are interested in solving
vectorial partial differential equations (PDE) employing spherical
components and spherical coordinates. The advantage of this choice is
the possibility to impose exact boundary conditions at the surface of
the star. The drawback, of course, are the coordinate singularities
appearing in the PDEs, but with spectral methods those difficulties
can be easily overcome if we introduce the angular potentials $ \eta
$ and $\mu$ defined as (\citealt{BM1990}, \citealt{BFG1996})
\be\label{angpote} B_\theta= \partial_{\theta}
\eta - \frac{1}{\sin \theta} \partial_\phi \mu, \; \; \;
B_\phi=\frac{1}{\sin \theta} \partial_\phi \eta +\partial_\theta \mu
\ee
or analogously
\be\label{angpotm}
 \triangle_{\theta, \phi} \eta= \frac{\partial^2 \eta}{\partial
\  \theta^2}
 +\frac{\cos \theta}{\sin \theta} \frac{\partial \eta}{\partial \theta} 
 +\frac{1}{\sin \theta^2} \frac{\partial \eta}{ \partial \phi^2}
\ee
$$=\frac{\partial B_\theta}{\partial \theta} 
+ \frac{\cos \theta}
{\sin  \theta} B_\theta +\frac{1}{\sin \theta}
 \frac{\partial
  B_\phi}{\partial \phi} $$
Now, if we expand $\eta$ in spherical harmonics we have
\be\label{sphrep}
 \triangle_{\theta, \phi}\eta=-l(l+1)\eta
\ee
Analogously for $\mu$
\be\label{angpotm2}
\triangle_{\theta,\phi} \mu =\partial_ \theta B_\phi +\frac{\cos
    \theta}{\sin \theta} B_\phi -\frac{1}{\sin \theta} \partial_\phi
    B_\theta
\ee

To illustrate the use of these potential by some examples, we shall
mention that the divergence of a vector $\vec{B}$ can be written
\be\label{diver} div \vec{B}=\partial_r B_r + \frac{2}{r} B_r
+\frac{1}{r} \left( \partial_\theta B_\theta +\frac{\cos \theta}{\sin
\theta} B_\theta + \frac{1}{\sin \theta} \partial_\phi B_\phi\right),
\ee
which reads, in spherical harmonic representation, \be\label{armrep}
\partial_r B_r +\frac{2}{r} B_r-\frac{1}{r} l(l+1) \eta\,. \ee
In the same way, to deal with an equation like
$\triangle\vec{B}=\vec{S}$ for a divergence-free vector $\vec{B}$, the
decomposition in potentials of both $\vec{B}$ and $\vec{S}$ leads to
solve \be\label{LapBr} \frac{d^2 B_r}{dr^2}+\frac{4}{r} \frac{d}{dr}
B_r + \frac{1}{r^2}\left(2-l(l+1)\right) B_r=S_r\,, \ee
\be\label{lape} \frac{d^2 \eta}{dr^2}+\frac{2}{r} \frac{d \eta}{dr}
+\frac{1}{r^2} (-l(l+1)\eta+2 B_r)) =\eta_S\,, \ee \be\label{Lapmu}
\frac{d^2 \mu}{dr^2} +\frac{2}{r} \frac{d \mu}{dr} -\frac{1}{r^2}
l(l+1) \mu =\mu_S\,, \ee where Eq.(\ref{LapBr}) was obtained using
Eq.(\ref{armrep}) and the divergence-free nature of $\vec{B}$. Notice
that the equation for the toroidal part $\mu$ is decoupled from the
other two equations.\\

 Having obtained ordinary differential equations for the radial
variable, we see that the problem of the singularities on the axis
$\theta=-\pi,~\theta=\pi$ is easily solved by means of spectral
decomposition on spherical harmonics. Similarly, problems of
singularities at $r=0$ are solved by expanding the solution on a
complete set of functions that have a good analytical behaviour at
$r=0$ (e.g. Chebyshev polynomials or a linear combination of them, see Appendix of Paper I).

\subsection{Divergence-free decomposition}

We shall now describe in more detail the new method that is used to
apply Helmholtz theorem and decompose a vector field into a
divergence-free and a potential parts. Consider again the equations of
motion Eq.~(\ref{MHDa},~\ref{MHDav}), where $S_i$ is a source term
that is supposed to be known at the time $t_{j-1}$, and that
we have to decompose into a divergence-free part
${\tilde S}_i$ and a potential part $\partial_i \Phi$, \be\label{decc}
S_i={\tilde S}_i+\partial_i \Phi\,. \ee
The ordinary way to make this decomposition is described in all analysis textbooks and was applied in Paper I: one simply takes the
divergence of the Eq.~(\ref{decc}) and solves the Poisson equation
\be\label{dec22} \triangle \Phi= div S\,.\ee

 Once $\Phi$ is obtained\footnote{As already mentioned, $\Phi$ is defined
within a harmonic functions $\Phi_A$ that is kept as an additional
degree of freedom used for the implementation of BC.}, we got ${\tilde S_i} $ by ${\tilde
S}_i=S_i-\partial_i \Phi~.$\\

The problem in the numerical implementation of this method is that in
order to obtain ${\tilde S}_i$ from $\Phi$ one needs to compute its gradient, while the numerical computation of a derivative
means higher round-off errors and generation of numerical
instabilities. To avoid this problem, in the MHD code, we directly get
$\tilde{S}$ by first computing ${\cal
  C}_\eta$, the $\eta$ angular potential of the curl of $S$, and then solving the system
\be\label{dec1} (curl {\vec {\tilde S}})_\eta=(curl {\vec S})_\eta, \;
div {\vec{\tilde S}} =0\,. \ee
After an expansion in spherical harmonics, this system reads
\be\label{dec2}
 \frac{d \eta}{dr}+\frac{\eta}{r}-\frac{{\tilde S}_r}{r}=
{\cal C}_\eta, \; \;  
 \frac {d {\tilde S_r}}{dr} +\frac{2}{r} {\tilde S}_r - \frac{l(l+1)}{r}\eta=0\,,
\ee
where $\eta$ is the potential of $\tilde{S}$. An expansion in Chebyshev
polynomials leads to an algebraic linear system whose matrix is a $2
(N_r-1)\times(2N_r-1)$ one ($N_r$ being the number of coefficients in the
expansion) that can be reduced easily to a 8-diagonals matrix, and in this way,
$\tilde{S}$ is solved without using $\Phi$.

\subsection{Solution for the equation of a divergence-free vector}

In the general case, the MHD equations are not linear and the solutions with
zero viscosity (infinite Reynolds number) or/and zero conductivity (infinite
magnetic Reynolds number) present discontinuity in the derivatives (infinite
shear). Consequently, they cannot be described by numerical solutions with
finite resolution and require specific techniques. Hence, in the following, we
mainly consider the general case with both conductivity $\sigma$ and viscosity
$\zeta$ not equal to zero, particular cases for which $\zeta$ or/and $\sigma$
vanish ({\it i.e.} linearized problems) being only partially discussed in the
next Section, but mainly kept for another article. Furthermore, we describe
here only the method used to solve the equation for the magnetic field $\vec{B}$,
the case of the equation of the density momentum $\vec{{\cal P}}$ being
similar since the dispersive equation for the magnetic field $B$
[Eq.~(\ref{Cns})] can be written \be\label{B} \partial_t \vec{B} + 4 \pi c^2
curl(\sigma \, curl
\vec{B})= S\,, \ee where $S$ contains all other terms.\\

To proceed, we use the decomposition of the vector $\vec{B}$ in radial
component $B_r$ and two angular potential $\eta$ and $\mu$ as
explained above (Eq.~\ref{angpote}). The source term is calculated at time
$t_{j+1/2}$ using the previous values at time $t_j$ and $t_{j-1}$ by
\be \label{exeq}
S^{j+1/2}=\frac{3\,S^j-S^{j-1}}{2}\,.
\ee

 For the moment, we will consider $\sigma$ constant, the non-constant
 case being technically identical (see Appendix of Paper I). Using the
 identity
$$ curl\,curl\,\vec{B}=-\triangle\, \vec{B} +\,\overrightarrow{grad} \left(div \vec{B}\right)$$ 
and taking into account the divergence-free nature of $\vec{B}$, an expansion in spherical
harmonics gives the system [see Eqs.~(\ref{LapBr},\ref{lape},\ref{Lapmu})]
\be\label{Breq}
\partial_t B_r + \sigma \left[\frac{d^2 B_r}{dr^2}+\frac{4}{r} \frac{dB_r}{dr}
+\frac{1}{r^2}(2-l(l+1))B_r \right ]=S_r
\ee
\be\label{etaeq}
\partial_t \eta +\sigma \left[\frac{d^2 \eta}{dr^2}+\frac{2}{r} 
\frac{d\eta}{dr}+\frac{1}{r^2}(-l(l+1)\eta +2 B_r)] \right ]=S_{\eta}
\ee
\be\label{mueq}
\partial_t \mu+\sigma \left [\frac{d^2 \mu}{dr^2}+\frac{2}{r} \frac{d
    \mu}{dr}-\frac{1}{r^2}l(l+1)\mu \right ]=S_{\mu}
\ee
where $S_\eta$ and $s_\mu$ are respectively the angular potentials of
the source $S_i$. Remember that the condition $div\vec{B}=0$ reads
\be\label{div}
\frac{d B_r}{dr}+\frac{2}{r} B_r -\frac{1}{r} l(l+1) \eta=0\,.
\ee
Notice also that the equations for $B_r$ and $\mu$ are decoupled and behave like  
the Poisson equation.\\

There are different strategies to solve the above system: we can, for
example solve the Eq.~(\ref{Breq}) and compute $\eta$ by using the
divergence [Eq.~(\ref{div})]:
$$ \eta=\frac{1}{l(l+1)}\left(r \frac{d B_r}{dr}+2 B_r\right).$$ This
 method is very simple, but its drawback is that we have to perform a
 derivative with respect to $r$ of $B_r$, that generates high spacial
 frequency noise, which in turn produces numerical instabilities in a
 MHD code. We can also use the method described in Paper I, but
 we present here a different one that is more robust.\\

This new method is the following: After the time discretisation, a second order (in time) implicit
 scheme consists in re-writing the Eqs.~(\ref{Breq},~\ref{etaeq}) in
 the following way:
 \beqn \label{Breqd}
B_r^{j+1}-\frac{\delta t}{2} \sigma \left [\frac{d^2 B_r^{j+1}}{dr^2}+\frac{4}{r} \frac{d
 B_r^{j+1}}{dr} +\frac{1}{r^2}(2-l(l+1)) B_r^{j+1} \right ]=\nonumber\\
B_r^j+\frac{\delta}{2} \sigma \left[\frac{d^2 B_r^j}{dr^2} +\frac{4}{r} \frac{d
    B_r^j}{dr} +\frac{1}{r^2}(2-l(l+1)) B_r^j \right] +\delta t\, S_r^{j+1/2}\,,
\eeqn

\beqn\label{etaqed}
\eta^{j+1}-\frac{\delta t}{2} \sigma \left [ \frac{d^2 \eta^{j+1}}{dr^2}
  +\frac{2}{r} \frac{d \eta^{j+1}}{dr} +\frac{1}{r^2}( 2 B_r^{j+1}
  -l(l+1)  \eta^{j+1})\right ]=\nonumber\\
\eta^{j} +\frac{\delta t}{2} \sigma \left[\frac{d^2 \eta^j}{dr^2} +
  \frac{2}{r} \frac{d \eta^j}{dr} +\frac{1}{r^2}(2 B_r^j-l(l+1)  \eta^j)
\right ]+\delta t\, S_\eta^{j+1/2}\,,
\eeqn
where $\delta t$ is the time increment and $S^{j+1/2} $ is the value
of the source extrapolated at time $ t_j +\delta t/2$ as explained by Eq.(\ref{exeq}).
Note the presence of the singular term $(2 B_r^{j}-l(l+1)\eta^j)/r^2$
at the RHS of the Eq.~(\ref{etaqed}). An elementary study of the
analytical properties of the above system of equations shows that the
solutions vanish at least as $r^{l-1}$ at the origin
$r=0$.\footnote{The value $l=0$ is forbidden, because we treat a pure
vector field.}  Consequently for $l \ge 3$, the solutions vanish fast
enough to avoid actual singularities in the apparently singular terms
of Eq.~(\ref{etaqed}) RHS. For $l=1,2 $, those terms are indeed
singular, but the solution is not singular because they do compensate
exactly. However, in order to overcome this numerical difficulty, we
re-write Eq.~(\ref{etaqed}) by using Eq.~(\ref{div}) as
follows:
\beqn\label{etaqed1} \eta^{j+1} -\frac{\delta t}{2} \sigma \left [\frac{d^2
    \eta^{j+1}}{dr^2}+\frac{2}{r} \frac{d \eta^{j+1}}{d
    r}-\frac{1}{r^2}(l(l-1)  \eta^{j+1}) \right ] =\nonumber\\
\eta^{j}+\frac{\delta t}{2} \sigma \left [ \frac{d^2 \eta^j}{d^2r}
  +\frac{2}{r} \frac{d \eta^j}{dr} -\frac{1}{r^2}l(l-1) \eta^j \right ]
+(S^1+S_{\eta}^{j+1/2})\, \delta t\,, \eeqn where $S^1$ is a new source term:
\be\label{S1} S^1=\frac{\sigma}{2r^2} \left[ -\frac{2}{l+1}\left(r \frac{d
      B_r^{j+1/2}}{dr} +2B_r^{j+1/2}\right) +2 B_r^{j+1/2} \right] \ee It easy
to see that for $l=1$ or $l=2$, since $B^l_r=b_lr^{l-1}+...$ and $\eta^l=e_l
r^{l-1}+...$, all the terms in Eq.~(\ref{etaqed1}) are regular.\\

Then, the way to proceed consists in first, getting homogeneous solutions by doing the following:\\
\\
a) - Compute an homogeneous solution of the system of Eq.~(\ref{Breqd},~\ref{etaqed}): solve with $S_r^{j+1/2}=0$ for calculating $B_r^{j+1}$, let $B_r^h$ be this solution.\\ 
b) - Compute an homogeneous solution $\eta_h$ of the Eq.~(\ref{etaqed1}): solve with $S^1=0$ and $S_{\eta}^{j+1/2}=0$\\
c) - Compute a particular solution $\eta^{h1}$ of the
Eq.~(\ref{etaqed1}) using $S^{j+1/2}=0$ and $S^1$, the latter being computed from $B_r^h $\\  
d) - Impose, by using the homogeneous solution $\eta_h$, that the
divergence given by the Eq.~(\ref{div}) vanishes at the boundary ($r=R$).\\

 In such a way, we have a homogeneous solution of the system
(\ref{Breqd}, \ref{etaqed}) that satisfies exactly the condition of vanishing divergence ($S_r^{j+1/2}=0$, $S_\eta^{j+1/2}=0$) at
the boundary. All those homogeneous solutions are stored and have
to be computed at each time-step whether $\delta t\,\sigma$ depends on
time. The complete solution is computed by obtaining, in a similar way,
a particular solution and by using the homogeneous solution to satisfy
the BC. The condition $div \vec{B}=0$, as was already said, is
fulfilled exactly by construction at the boundary and approximately in
the interval $0 \le r \le R$. For a given number
of degrees of freedom $N_r$ in $r$, the error $\epsilon$ depends on the value of $\delta
t\,\sigma$. In practical cases, we had $10^{-11} \le \epsilon \le
10^{-5}$.
 
\section{Non-dissipative MHD}
\label{non_dissipative_mhd}

In the previous section, we have discussed the general MHD problem
(non-vanishing viscosity and resistivity). As explained, in this case, the
presence of viscosity and resistivity allows us to easily impose BCs and help
to stabilize. Yet, there are astrophysical steady-state problems for which we
really need a non-dissipative code, for example while probing the stability of
hydromagnetic configurations. For such a class of problems, linearized MHD has
to be used, viscosity and resistivity being no more necessary to smooth the
solutions, being on the contrary to be avoided. Hence, the
implementation of BCs has to be done in a different way. This technology is in
progress, and as a first illustration, we present here an example of application inspired by a model of
$\gamma$-ray bursts proposed recently by \cite{PH2005} and based on previous
proposals by
\cite{kr98} and \cite{dl98}.\\

This model consists in a just born differentially rotating and
magnetized NS, that during the cooling time (and consequently the
shrinking time) undergoes a phase transition and becomes a
{\it differentially rotating stratified quark star}. At this point, the
poloidal magnetic field is stretched by the differential rotation and
a fraction of the rotational kinetic energy related with the
differential rotation is transformed into magnetic energy of a
toroidal field. Orders of magnitude estimation shows that for a star
initially rotating at $200$ Hz, with an amount of some few percent of
the total rotational kinetic energy due to the differential rotation,
energy budget is met. Moreover, a strong toroidal magnetic field ($
\sim 10^{15}$ gauss) is generated. If there are no losses, the
magnetic field reaches a maximum and starts to decrease, transforming
its magnetic energy back into differential kinetic energy (at constant
kinetic momentum). The duration of this twisting period depends on the
intensity of the preexisting poloidal magnetic counterpart $\vec{B}_p$
($\sim 100$ rotation periods for $B_p=10^{13}$ gauss). From a
numerical point of view, the winding time of the toroidal magnetic
field $B_{phi}$, takes about hundred rotation periods of the star, meaning
that it is crucial to avoid important energy losses.\\

The low dissipation condition cannot be fulfilled with the code
described above reasonable numerical viscosity and
resistivity. Hence, we modified it, taking into account the
particularity of the problem in order to eliminate numerical
dissipation. The strong stratification of the star suggests the
approximation $V_r=0$ and $V_\theta=0$, since motions of the
matter inside the star tend to be parallel to isopotential surfaces: $V_r\sim 0$, $V_\theta \sim
0$. Consequently, only evolutions of the toroidal components of the
magnetic field $B_\phi$ and of the velocity $V_\phi$ are pertinent to
our problem. Moreover, as a first step, we consider in the following an
incompressible fluid and Newtonian gravity.

\subsection{Evolution of the toroidal component $B_\phi$}

The equations of motion for the toy model are
\be
\partial_t \vec{B}\,=\,\textrm{curl}\left(\vec{V}\,\wedge \vec{B}\right)\,,
\ee
and
\be
\partial_t
\vec{V}\,=\,\frac{1}{4\pi\,n}\,\left(\textrm{curl}\vec{B}\right) \wedge \vec{B}\,.
\ee

Taking into account the assumptions described above, the axisymmetry and the
equatorial symmetry, the equations to be solved read
\be\label{Bfiev}
\partial_t B_\phi-B_r\left(\partial_r V_\phi-\frac{1}{r} V_\phi\right) -\,\frac{B_\theta}{r}
\left(\partial_\theta V_\phi-\frac{1}{r} \frac{\cos \theta}{\sin \theta}
V_\phi\right)=0\,,
\ee
\be\label{Vphiev}
 \partial_t V_\phi-\frac{1}{4\pi\,n} \left[ B_r\left( \partial_r B_\phi+\frac{1}{r} B_\phi\right)  
   +\frac{B_\theta}{r}\left(\partial_\theta B_\phi+ 
\frac{\cos \theta}{\sin \theta} B_\phi\right) \right] =0\,,
\ee
from which we see that in the case of rigid rotation, $V_\phi=\Omega\,r\,\sin \theta$, $B_\phi$ does not evolve.\\
 
In what follows, we shall assume in addition that the poloidal
magnetic field can be written (taking into account the symmetries of
the problem) $$B_r\,=\,\cos \theta\,(b1+b_2\, r^2 \cos(2 \theta)+...)\,,$$ $B_\theta$ being obtained by the condition $\textrm{div} \vec{B}=0$.\\

 We have to impose BCs in order to have an unique solution. A study of Eqs.(\ref{Bfiev}) and (\ref{Vphiev}) shows that this system is
hyperbolic, and consequently, one BC can be imposed at the surface of the
star. Physically, what occurs is that a radial Alfv{\'e}n wave is created in order to satisfy
the given BC. Since in the local approximation, the phase velocity $V_{Al}$ of
an Alfv{\'e}n wave verifies
\be
V_{Al}^2\,=\,\frac{w^2}{k^2}\,=\frac{\left(\vec{k}\cdot\vec{B}\right)^2}{4
\pi n\,k^2}\,,
\ee
for an Alfv{\'e}n wave propagating
radially, the phase velocity is \be\label{Alf} V_{Al}^2=\frac{1}{4 \pi
n} B_r^2\,, \ee which is maximum at the pole ($\theta=0$) and vanishes
for $\theta=\pi/2$ (the equator), meaning some possible numerical troubles due
to the large variation of the timescale.\\

Hence, the technique to solve this system is to work in configuration space
in $\theta$ and in the Chebyshev space for $r$. Once the
discretisation of $\theta$ is done, we are confronted with $N_\theta$
hyperbolic systems that can be solved implicitly. Since the Alfv{\'e}n
velocity can be very small for $\theta$ close to $\pi/2$, the number of
degrees of freedom $N_r$ for $r$ must be high enough to describe the
Alfv{\'e}n wave that propagates very slowly along the radius. In
conclusion, $N_r$ depends on $N_\theta$. Note that for $\theta=\pi/2$
(the equator) the system is completely degenerated, and $B_\phi$
vanishes as it is anti-symmetric with respect to the equator.\\

To conclude, we emphasize that the finale solution of the system strongly depends on the chosen BCs. For
example, if we choose $B_\phi=0 \mid_{r=R}$, the Poynting vector
vanishes at the surface, and the energy is conserved\footnote{Note that the radial
component $J_r$ of the current vanishes at the surface too:
$$ J_r=\frac{1}{r}( \partial_\theta B_\phi +
\frac{\cos \theta}{\sin \theta} B_\phi )\,.$$}. Thus, in the next Section, we
shall provide some examples with conditions that allows us to easily check the energy conservation and consequently to
test the accuracy of the code. More precisely, we shall use for the
non-dissipative case the condition $B_r=0 \mid_{r=R}$ at $t=0$, which makes
the operator degenerated at the surface, implying that no BC is required but
the Poynting vector does vanish at the surface.

\section{Tests}
\label{tests}

In the following, we shall illustrate the stability of the code, but mainly
with tests that concern linear (or linearised) problems. Yet, we would like to point
out again that linear problems, when using spectral methods, are the most
severe ones. Indeed, for a linear problem, a spectral code has to work with vanishing viscosity and
conductivity, and since spectral methods have no intrinsic numerical
dissipation, numerical instabilities have greater
opportunity to grow than in the non-linear case, where viscosity and
resistivity can be increased until the code becomes stable. This is the reason
why the stability conditions illustrated in the following were obtained in a non-trivial
way, that will nevertheless be explained in detail somewhere else.

\subsection{Dissipative case}
\begin{figure}[h]
\begin{center}
\includegraphics[width=0.6\columnwidth,clip]{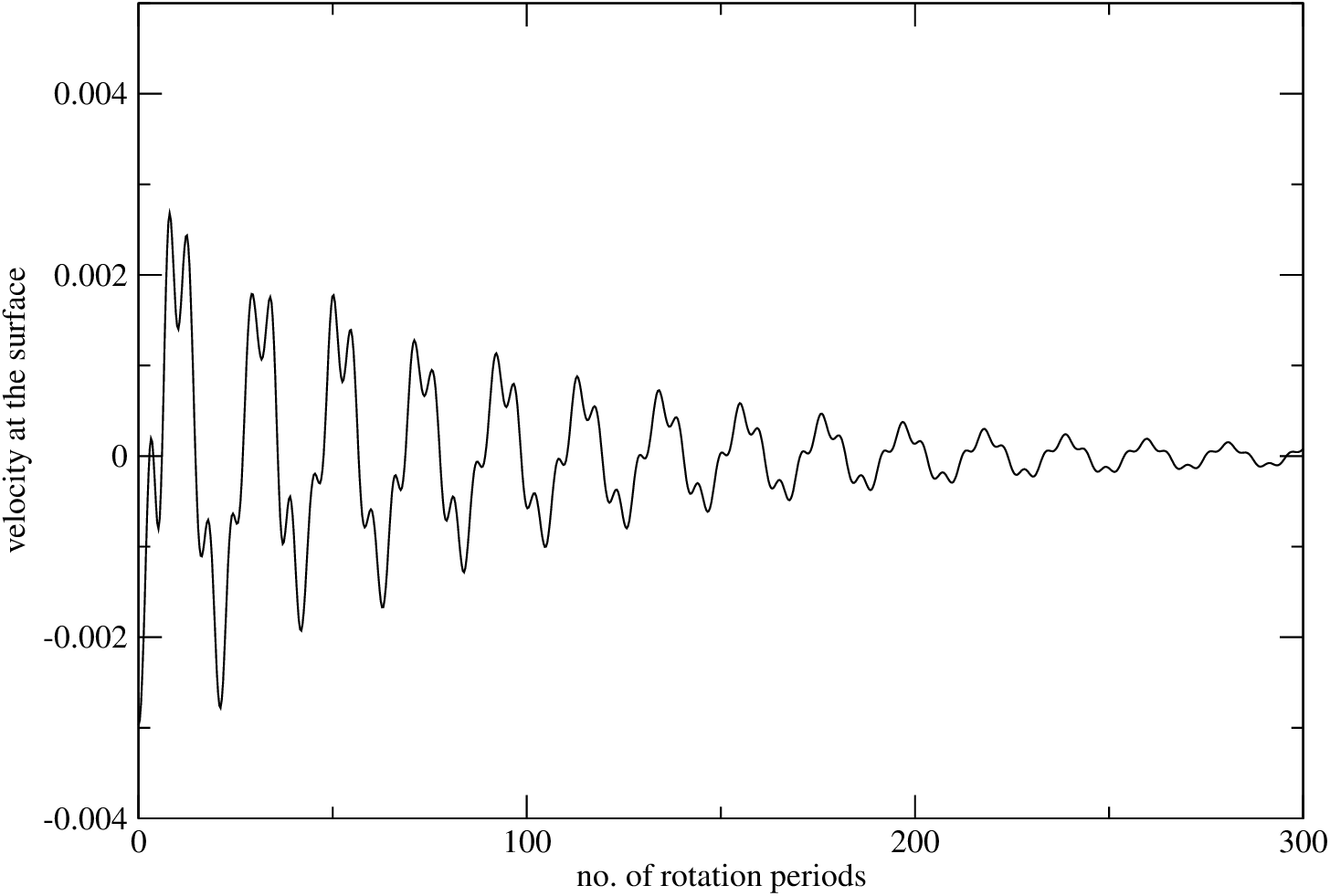}
\caption{Time evolution of the velocity $v_\phi$ of a given point on the surface of the star for
  a dissipative code with excited $l=2$ r- and f-modes.}
\label{wavedissips}
\end{center}
\end{figure}

Since the dissipative case is far being the most interesting one from the
numerical point of view, we only provide one example of evolution 
 in the case of a hydrodynamical coupling between an r-mode 
(lower frequency) and $l=2$ f-mode (higher frequency).
The calculation pictured on Fig.\ref{wavedissips} shows the time evolution of the component
$v_\phi$ of the velocity on the surface of the star.
The calculation is done with the number 
of coefficient in spectral expansion equal to 24, 6 and 4 in $r$, $\theta$ 
and $\phi$
directions, respectively, and the time-step is $10^{-2}$ (in the
computational units in which the period of rotation is equal unity). The 
dimensionless viscosity parameter equals 0.001. 

\subsection{Non-dissipative case}
\begin{figure}[h]
\begin{center}
\includegraphics[width=0.6\columnwidth,clip]{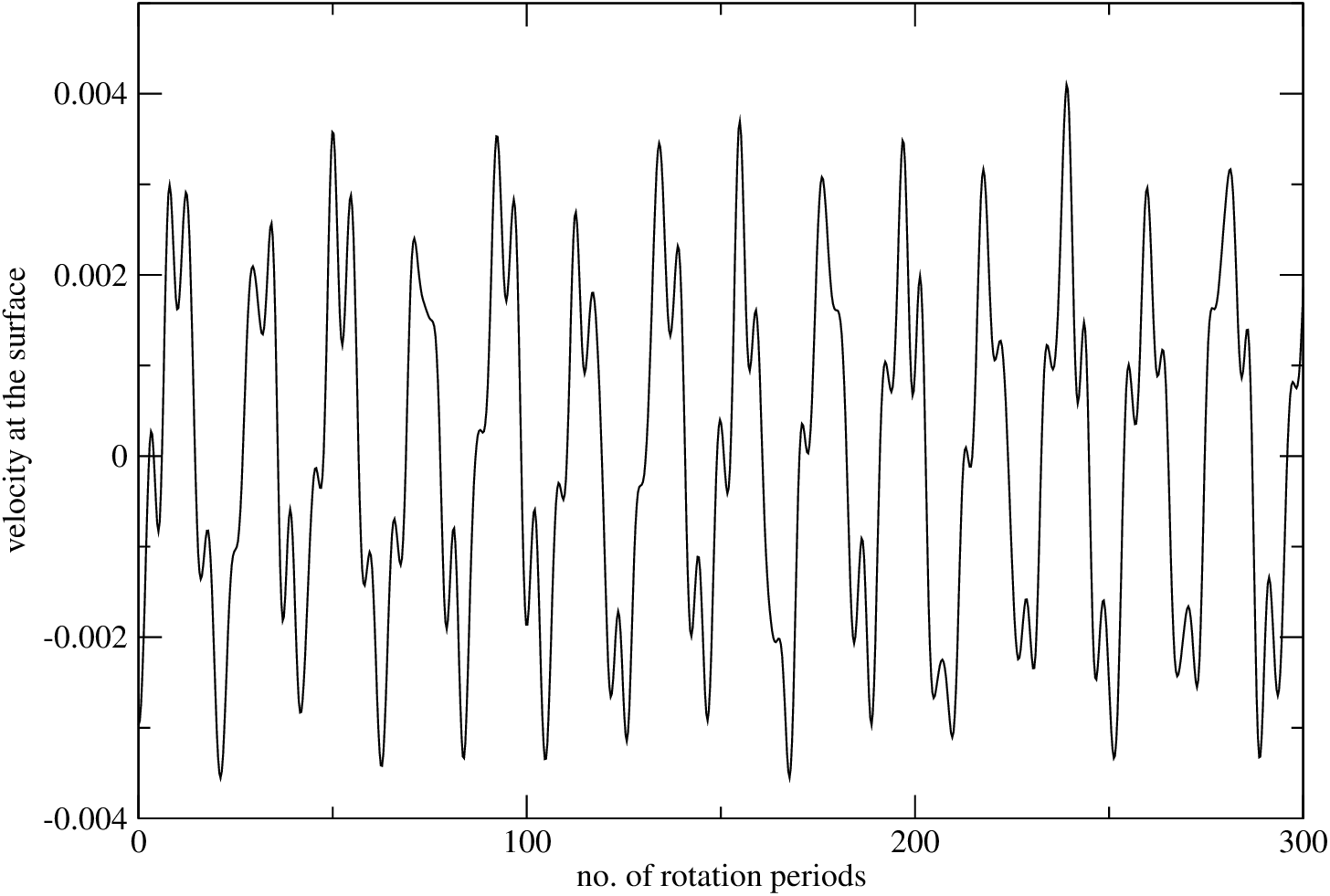}
\caption{$v_\phi$ component of the velocity for a given point on the surface of the star 
with excited $l=2$ r- and f-modes, versus time.}
\label{wave8000s}
\end{center}
\end{figure}

\begin{figure}[h]
\begin{center}
\includegraphics[width=0.6\columnwidth,clip]{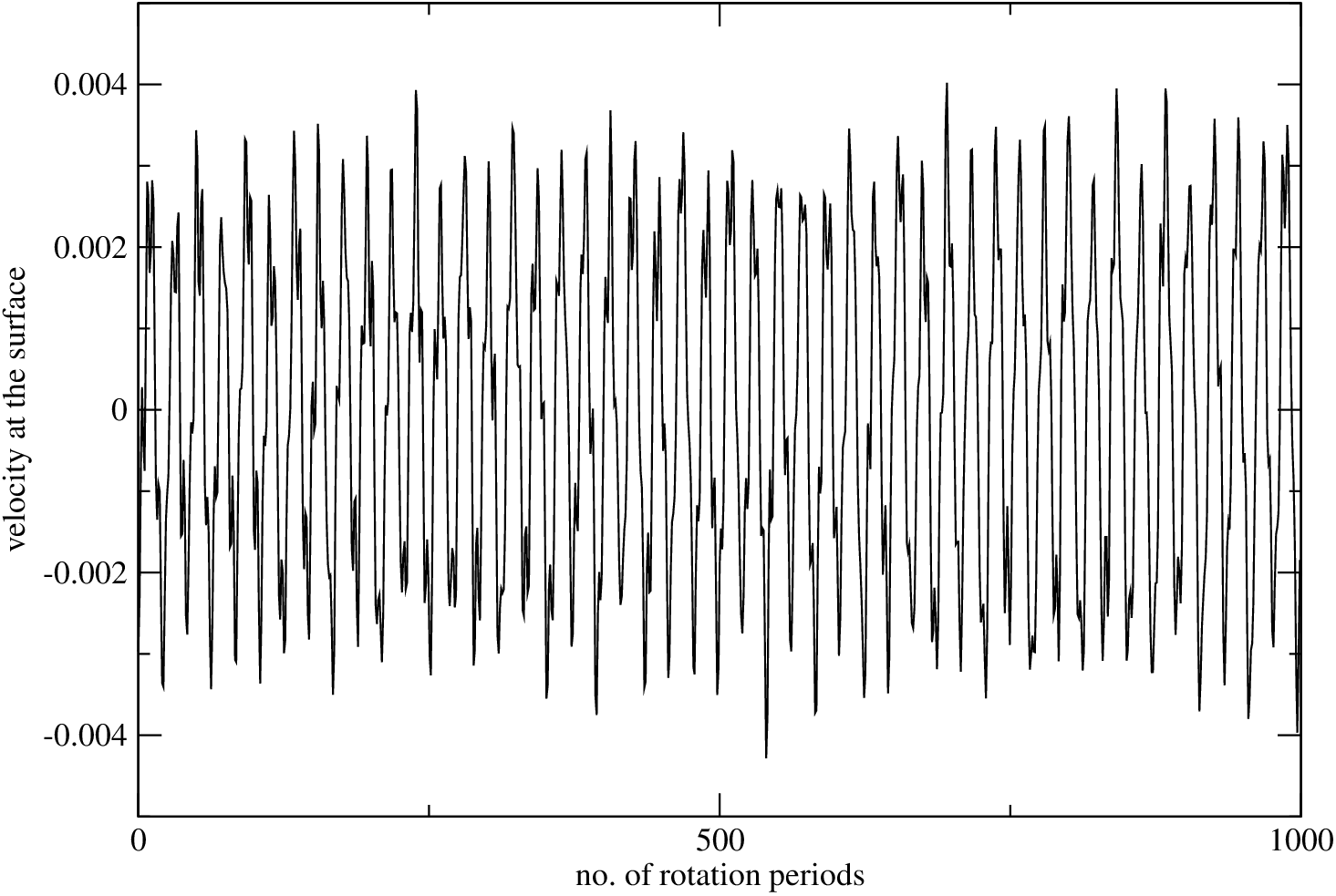}
\caption{The same as on Fig.~\ref{wave8000s}, but for a longer run, in order 
to illustrate the lack of dissipation and the long-term stability of the code.}
\label{wave20000s}
\end{center}
\end{figure}

We also present a couple of illustrative tests of the spectral code, that were
performed with the linear version, neglecting all dissipative terms. On
Figs.~\ref{wave8000s} and \ref{wave20000s}, the case 
of a pure hydrodynamical coupling between an r-mode and $l=2$ f-mode is shown.  
As in previous subsection, the number of coefficient
in spectral expansion is again 24, 6 and 4 in $r$, $\theta$ and $\phi$
directions, respectively, while the time-step is $10^{-2}$.\\

\begin{figure}[h]
\begin{center}
\includegraphics[width=0.8\columnwidth,clip]{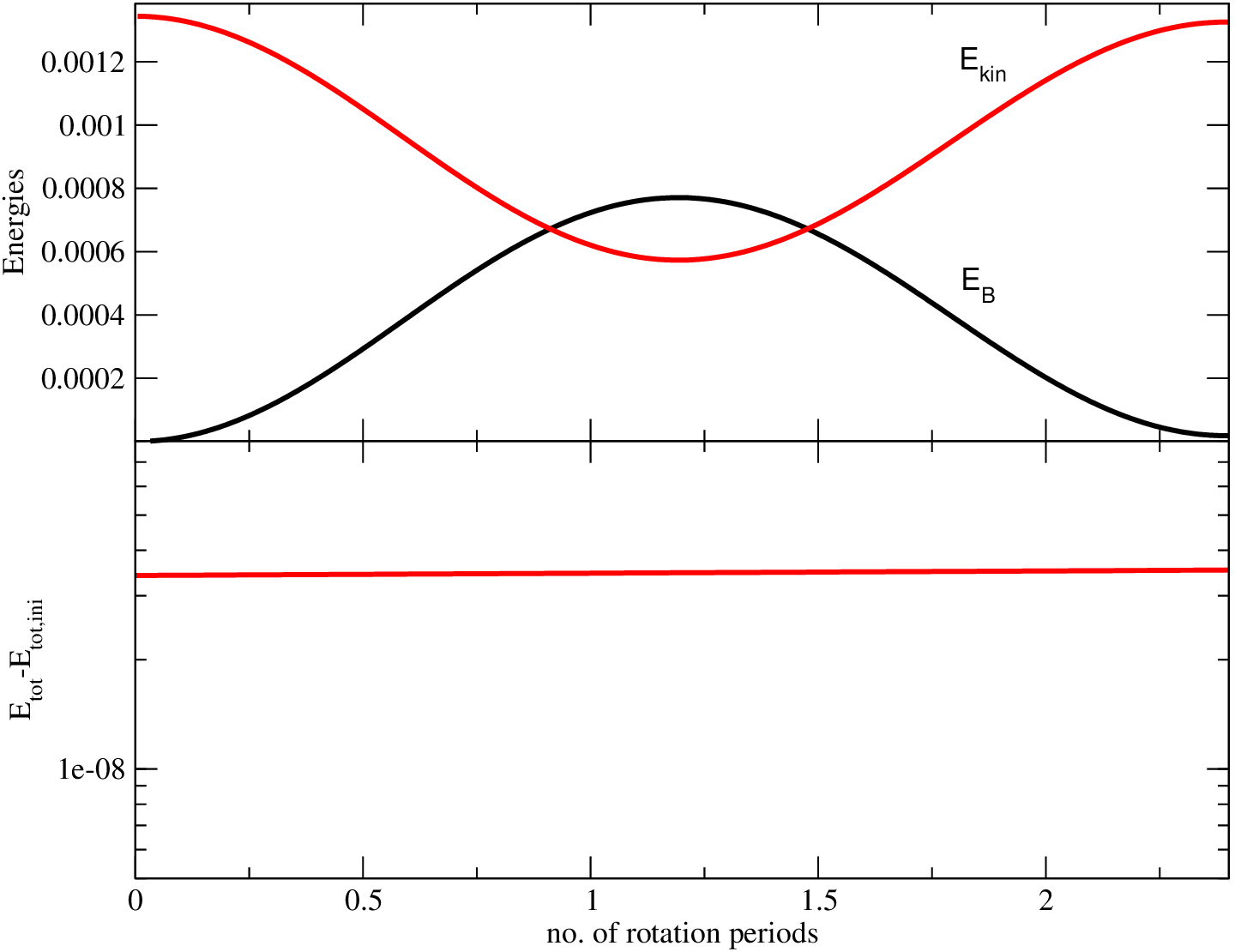}
\caption{The interplay between the kinetic energy $E_{\rm kin}$ and 
magnetic energy $E_{\rm B}$ as well as the variation of  
the total energy during the creation of a toroidal component of the
magnetic field from an initial purely poloidal magnetic field, due 
to differential rotation (see text for details).}\label{ekin_eb}

\end{center}
\end{figure}

Last but not least, we shall also plot the conservation of energy in a linear
MHD code using the no-radial-magnetic field condition (see end of previous
Section) and no dissipation term. The chosen model is axisymmetric (24
coefficients in $r$-direction and 16 in $\theta$-direction) with the time-step equal $10^{-3}$ and with an initial
$v_\phi$ profile $r\sin\theta-r^3\sin^3\theta$ and an initial magnetic field
with only a radial component: $B_r=\sin\theta(1-(r/R)^2)$. On
Fig.~\ref{ekin_eb}, it can be seen that for 2.5 rotation periods (2500 time-steps), energy is
conserved up to the value of $10^{-8}$ (the difference between the total
initial and final energy), meaning a relative variation around $10^{-5}$ only.
As expected, at the end of the cycle, the star has gained strong differential
rotation and a poloidal component of the magnetic field.\\

\section{Conclusions}
\label{conclusions} 
We have described a code based on spectral methods able to handle 3D
relativistic MHD problems in a sphere (e.g. the interior of a slowly rotating
neutron star). The code works with spherical coordinates and spherical components of
the velocity $\vec{V}$ and of the magnetic field $\vec{B}$. The use of
spherical coordinates and spherical components allows us to impose exact BCs
on the surface of the star. Moreover, thanks to spectral methods, the
singularities present on the $z$ axis and at $r=0$, which result from
spherical coordinates, can be treated properly. We want to point out that the
numerical solution is obtained in the {\it whole } star, while numerous
dynamical spectral codes (for instance in geophysics) are restricted to some shells.\\ 

As far as physics is concerned, order of magnitude considerations show that
for most of the known compact stars, the acoustic time scale is much shorter
than other timescales (Alfv{\'e}n, Brunt-V\"ais\"al\"a, rotation
timescale, ...). In order to easily have an efficient code, the
anelastic approximation, which filters acoustic waves, is employed. In such a way,
slow evolution of the star can be followed during tenths of its rotations.
Moreover, due to the intrinsic accuracy and low dissipativity of spectral
methods, the code is well suited to study the stability of steady-state MHD
configurations and other physical problems that we shall investigate elsewhere.

\subsection*{Acknowledgements}
\label{acknowledgements}
One of us (SB) thanks Dr. J. Novak for helpful discussions. 
LV and MB were supported, respectively, by the Marie Curie 
Intra-european Fellowships MEIF-CT-2005-025498 and MEIF-CT-2005-023644, 
within the 6th European Community Framework Programme.

{}

\begin{thebibliography}{}


\bibitem[Ant\'on {\it et al.}(2006)]{ant06}
Ant\'on, L., Anett, O., Miralles, J.A., Mart\'\i, J.M., Ib\'a\~nez, J.M.,
Font, J.A., Pons, J,A., ApJ {\bf 637}, 296 (2006) 

\bibitem[Bonazzola {\it et al.}(1996)]{BFG1996}
Bonazzola, S., Frieben, J., Gourgoulhon, E. Astrophys. J. {\bf
  3012},675 (1996)

\bibitem[Bonazzola \& Marck(1990)]{BM1990}
Bonazzola, S. Marck, J.A. J.Comp. Phys. {\bf 87},201 (1990)

\bibitem[Bucciantini {\it et al.}(2006)]{buc06}
Bucciantini, N., Thompson, T.A., Arons, J., Quataert, E., Del Zanna, L., Mon.\ Not.\ Roy.\ Astron.\ Soc.\ {\bf 368}, 1717 (2006)

\bibitem[Dai \& Lu(1998)]{dl98}
Dai, Z.G., Lu, T., Phys. Rev. Let. {\bf 81}, 4301 (1998)

\bibitem[De Villiers \& Hawley(2003)]{dvh03}
De Villiers, J.-P. Hawley, J.~F., ApJ {\bf 589}, 458 (2003)

\bibitem[Duez {\it et al.}(2005)]{du05}
Duez, M.~D., Liu, Y.~T., Shapiro, S.~L., Stephens, B.~C., Phys. Rev. D {\bf
  72}, 024028 (2005)

\bibitem[Gammie {\it et al.}(2003)]{gam03}
Gammie, J.C., McKinney, J.C., T\'oth, G., ApJ {\bf 589}, 444 (2003)

\bibitem[Goldreich \& Reisenegger(1992)]{gr92}
Goldreich, P., Reisenegger, A., ApJ {\bf 395}, 250 (1992)

\bibitem[Klu\'zniak \& Ruderman(1998)]{kr98}
Klu\'zniak, W., Ruderman, M., ApJ {\bf 505}, L113 (1998)

\bibitem[Koide {\it et al.}(1999)]{koi99}
Koide, S., Shibata, K., Kudoh, T., ApJ {\bf 522}, 727 (1999)

\bibitem[Komissarov(2004)]{kom04}
Komissarov, S.S., Mon.\ Not.\ Roy.\ Astron.\ Soc.\ {\bf 350}, 1431 (2004)

\bibitem[Lichnerowicz(1967)]{lich67}
Lichnerowicz, A., Relativistic Hydrodynamics and Magnetohydrodynamics, New York: Benjamin, (1967)

\bibitem[Mizuno {\it et al.}(2006)]{miz06}     
Mizuno, Y., Nishikawa, K.I., Koide, S., Hardee, P., Fishman, G.J., astro-ph/0609004

\bibitem[Paczy{\'n}ski \& Haensel(2005)]{PH2005}
Paczy{\'n}ski, B., Haensel, P. Mon.\ Not.\ Roy.\ Astron.\ Soc.\ Lett.\  {\bf 362}, L4 (2005) 

\bibitem[Shibata \& Sekiguchi(2005)]{shi05}
Shibata, M., Sekiguchi, Y.-I., Phys. Rev. D {\bf 72}, 044014 (2005)

\bibitem[Villain \& Bonazzola(2002)]{VB2002}
Villain, L., Bonazzola, S. Phys. Rev. D {\bf 66}, 123001 (2002) 

\bibitem[Villain, Bonazzola \& Haensel(2005)]{vbh05}
Villain, L., Bonazzola, S., Haensel, P. Phys.Rev. D {\bf 71}, 083001 (2005)

\bibitem[Wilson(1972)]{w72}
Wilson, J. R. ApJ {\bf 172}, 431 (1972)

\end{thebibliography}
\end{document}